# Low-Energy Boundary-State Emergence and Delocalization in Finite-sized Mosaic Wannier-Stark Lattices

Yi Kang, [1, §] Zhenjia Yu, [2, §] Xiumei Wang, [1, *] and Xingping Zhou [3, *]

[1] *College of Electronic and Optical Engineering, Nanjing University of Posts and Telecommunications, Nanjing 210003, China*

[2] *Portland Institute, Nanjing University of Posts and Telecommunications, Nanjing 210003, China*

[3] *Institute of Quantum Information and Technology, Nanjing University of Posts and Telecommunications, Nanjing 210003, China*

§ These authors contributed equally to this work.

[*]*Author to whom any correspondence should be addressed.*

[*]wxm@njupt.edu.cn

[*]zxp@njupt.edu.cn

The mosaic Wannier-Stark lattice has gained increasing prominence as a disorder-free system exhibiting unconventional localization behavior induced by spatially periodic Stark potentials. In the infinite-size limit, exact spectral analysis reveals an almost pure-point spectrum. There is no true mobility edge, except for ($M-1$) isolated extended states, which are accompanied by weakly localized modes with diverging localization lengths. Motivated by this spectral structure, we investigate the mosaic Wannier-Stark model under finite-size. In such systems, additional low-energy boundary-localized states emerge due to boundary residuals when the system length is not commensurate with the modulation period. These states are effectively distinguished and identified using the inverse participation ratio (IPR) and spatial expectation values. To explore their response to non-Hermitian perturbations, complex on-site potentials are introduced to simulate gain and loss. As the non-Hermitian strength increases, only the weakly localized states undergo progressive delocalization, exhibiting a smooth crossover from localization to spatial extension.

**Key words**

Mosaic Wannier-Stark lattices; Finite-size effect; Delocalization; Boundary-localized states;

The phenomenon of Anderson localization [1,2], originally describing the inhibition of wave propagation due to interference in disordered media, has been widely observed across various physical systems, from condensed matter [3-5] to optics [6-8], ultracold atoms [9-11], acoustics [12,13] and photonic lattices [14,15]. In three-dimensional (3D) environments with random disorder, such interference can induce a transition from extended to localized states at a finite threshold [16,17], giving rise to mobility edges [18-21]—energy values that mark this separation. Conversely, in one-dimensional (1D) settings with uncorrelated disorder, all eigenstates remain localized, and no such critical energy emerges [22,23]. Nonetheless, this limitation can be overcome by replacing randomness with deterministic modulations, such as those in quasiperiodic lattices. The Aubry-André-Harper (AAH) model [24,25] exemplifies this scenario, featuring a self-dual structure that results in an energy-independent localization transition. Extensions of this model, involving modified potentials [26], long-range couplings [27,28], and broken duality [29,30], have led to systems exhibiting energy-dependent mobility edges even in 1D geometries. Mosaic-type quasiperiodic lattices and flat-band configurations further enrich the range of such behavior. Empirical studies with platforms such as ultracold atomic gases [31,32], optical lattices [33-35], engineered photonic arrays[36], topolectrical circuits [37,38] and acoustic systems [39,40] have verified the existence of mobility edges in quasiperiodic structures. These developments have motivated the investigation of localization-transition models that lack both disorder and quasiperiodicity, such as lattices with spatially periodic Stark potentials [41,42].

In recent years, the mosaic Wannier-Stark lattice has attracted growing attention as a disorder-free platform featuring periodically modulated linear potentials, with the Stark field applied every $M$ sites [43-45]. Early theoretical investigations suggested the coexistence of localized and extended eigenstates, resulting in the claim of mobility-edge-like features despite the absence of disorder or quasiperiodic structure. However, subsequent exact spectral analysis showed that the spectrum becomes nearly pure point at the thermodynamic limit. All eigenstates are localized beyond exponential decay, except for $M-1$ isolated extended states. Around these extended states, a dense set of weakly localized states appears, characterized by diverging localization lengths. As a result, no rigorous mobility edge separates the localized and extended sectors in such systems[46]. Concurrently, non-Hermitian physics has seen substantial progress. A key focus has been on understanding the effects of engineered dissipation in open quantum systems [47-51]. It has been shown that non-Hermitian terms introduced via complex potentials can significantly reshape localization behavior, enabling the control of phase transitions and wave transport in both disordered and quasiperiodic environments [52-56]. In such contexts, gain and loss mechanisms have been found to stabilize localized phases or drive transitions across mobility edges [57-60]. The influence of non-Hermitian perturbations on disorder-free systems with pseudo-mobility-edge structures, including the mosaic Wannier-Stark lattice, remains insufficiently understood. In these systems, the combined impact of finite-size effects and non-Hermitian terms on localization warrants systematic investigation.

In this work, we investigate the spectral and localization properties of the mosaic Wannier-Stark lattice in finite systems with open boundary conditions. Unlike the idealized infinite model, a finite lattice does not always accommodate an integer number of modulation periods. When the system length $N$ is not divisible by the modulation period $M$, the residual $m = N \mod M$ sites at the two boundaries do not coincide with the positions where the periodic Stark potential is applied and therefore experience zero on-site potential. We show that this incommensurability leads to the emergence of $2m$ low-energy states whose wave functions are sharply localized near the edges. These boundary-localized states are clearly distinguishable from the weakly localized bulk states and the isolated extended states of the infinite system. To characterize these modes, we use the inverse participation ratio (IPR) [61] and spatial expectation values [62]. In the low-energy spectrum, the boundary states appear as isolated peaks in the IPR spectrum and are strongly localized near the lattice edges. In contrast, the bulk low-energy states exhibit either weak localization or extended behavior, showing a smooth variation of IPR with energy and remaining centered within the lattice. Furthermore, we introduce a non-Hermitian generalization of the model by adding an imaginary component to the Stark potential to simulate gain and loss. We find that the low-energy localized states undergo continuous delocalization as the non-Hermitian strength increases, reflected by suppressed IPR values and more uniform spatial profiles.

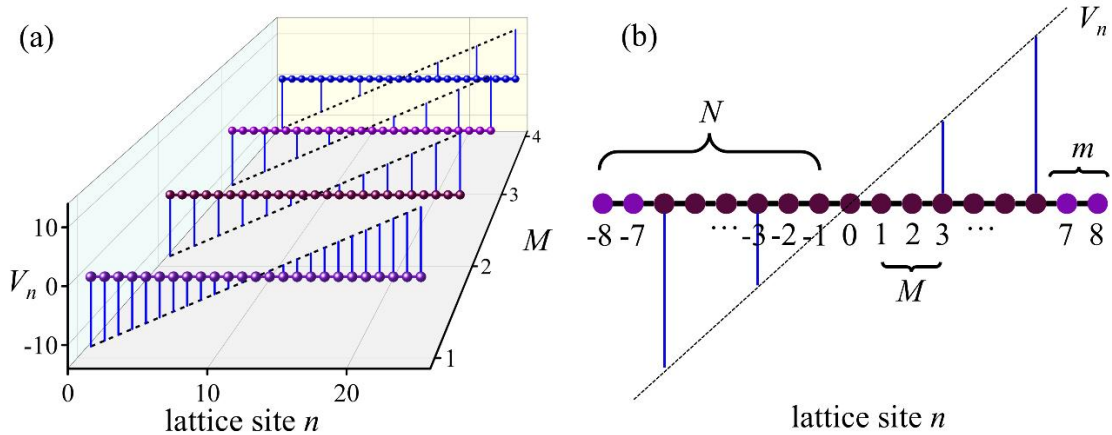

FIG. 1. Schematic of the mosaic Wannier-Stark lattice. A linear gradient potential $V_n = nF$ is applied at every $M$ lattice sites. (a) Configuration for different $M$. (b) Schematic of a finite lattice with $N$ sites per side and $m$ remainder sites.

We consider a disorder-free mosaic Wannier-Stark lattice [43-46] governed by the Hamiltonian:

$$H = \sum_n t(c_n^\dagger c_{n+1} + \text{H.c.}) + \sum_n V_n c_n^\dagger c_{n+1}, \tag{1}$$

where $c_n^\dagger$ and $c_n$ are the creation and annihilation operators at site $n$, respectively. The parameter $t$ denotes the nearest-neighbor hopping amplitude. The on-site potential $V_n$ represents a Stark field applied periodically every $M$ sites. Specifically,

$$V_n = \begin{cases} Fn, & n = 0, \pm M, \pm 2M, \pm 3M, ... \\ 0, & \text{otherwise}, \end{cases} \quad (2)$$

the parameter $F$ sets the strength of the applied Stark potential. The lattice is symmetric with respect to the center, with an equal number of sites on both sides [see Fig. 1(a)]. For a finite system, we define $N$ as the number of sites per side and introduce a remainder $m = N \bmod M$, which characterizes the residual segment due to incomplete periods of the Stark modulation [see Fig. 1(b)]. This residual incommensurability, which was not considered in Ref.[46], leads to different spectral features, as will be discussed below.

The energy spectrum of $H$ on an infinitely extended lattice, for $M \geq 2$, forms a pure point set. Most eigenfunctions exhibit localization stronger than exponential. However, there exist $(M-1)$-isolated energies corresponding to extended states, given by

$$E_\sigma = 2t \cos\left(\frac{\pi\sigma}{M}\right), \quad (3)$$

for $\sigma = 1, 2, ..., M-1$. The corresponding eigenfunctions are non-normalizable and take the form:

$$\psi_n^{(\sigma)} = \sin(n\pi\sigma/M). \quad (4)$$

The localized eigenstates are categorized into two branches: (i) high-energy states outside the band $(-2t, 2t)$, tightly localized near Stark potential sites, and (ii) low-energy states within the band, weakly localized but extending over many sites. To investigate the impact of remainder sites, we employ the IPR [61] and positional expectation analysis [62]. We use the IPR to differentiate between localized and extended states. For a wave

function normalized such that $\sum_{n=1}^{L}|\psi_n|^2=1$, the IPR is defined as:

$$\text{IPR}=\sum_{n=1}^{L}|\psi_n|^4. \tag{5}$$

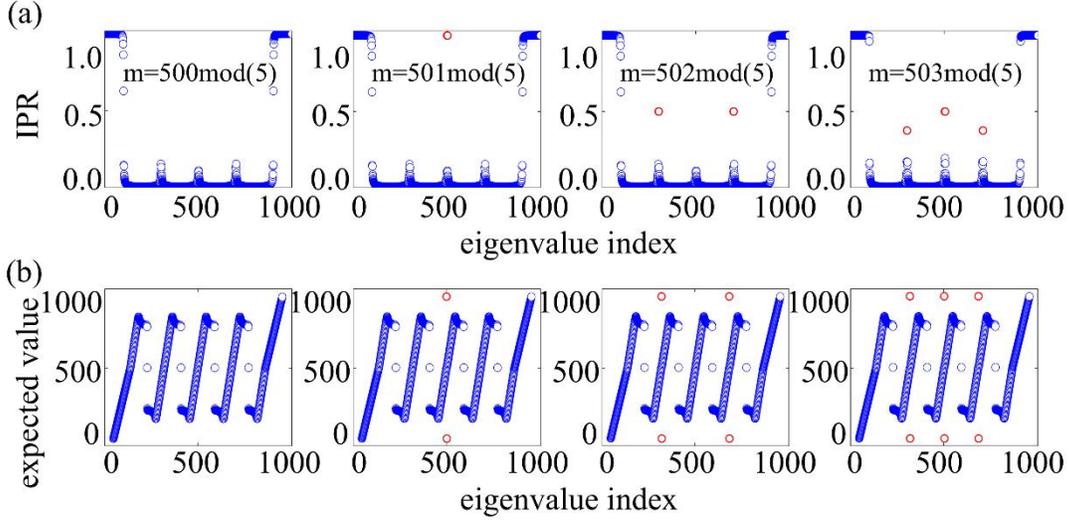

FIG. 2. (a) IPR of the eigenstates for $M=5$, $F/t=0.5$, with lattice sizes N=500 to 503, corresponding to remainders m=0 to 3. (b) Spatial expectation values of the corresponding eigenfunctions. For $M\neq 0$, boundary-localized states (indicated by red dots) appear at energies distinct from those of weakly localized states.

The overall IPR structure remains largely consistent across system sizes. Deviations are primarily reflected in several sharp peaks, which originate from boundary incommensurability due to residual sites, as illustrated in Fig. 2(a). In the high-energy regime, the IPR approaches unity, indicating strong localization. In contrast, smaller values in the low-energy regime suggest weaker localization. However, a set of large IPR values appears in the low-energy region when the remainder $m\neq 0$. Specifically, the $2m$ peaks emerge in the IPR spectrum, forming $m$ distinct pairs. Each pair of anomalies corresponds to a pair of boundary-localized states, one located at each end of the lattice. For $m = 0$, the low-energy wavefunctions correspond to weakly localized states. These states exhibit a higher degree of localization than the than exponential localization, yet they extend over a large portion of the lattice. Although some spatial concentration may occur, the wavefunctions remain confined to the central region and do not reach the boundaries. As such, they do not correspond to boundary-localized states. This behavior stands in sharp contrast to the regular eigenstate distribution observed in Ref.[46], where no low-energy boundary-localized features were reported. We introduce the spatial expectation value as a diagnostic measure to better resolve the extended states embedded within the low-energy band and clarify the structure of each boundary-state pair.

The spatial expectation values reflect the average position where a particle is likely to be found, and it is defined as follows:

$$\langle n \rangle = \sum_{n=1}^{L} n |\psi_n|^2. \tag{6}$$

The combination of IPR and spatial expectation values serves as an effective diagnostic for distinguishing extended, weakly localized, and boundary states, based on their intrinsic spectral and spatial characteristics. In the case of extended states, the wavefunctions are distributed across the entire lattice, with spatial expectation values centered near the middle. Weakly localized states tend to cluster near the center of the lattice, with spatial expectation values distributed within the bulk yet remaining away from the boundaries. However, the residual sites resulting from the incommensurate system length give rise to low-energy boundary-localized states, which are identified by spatial expectation values concentrated near the lattice edges (red dots).

As the energy of low-energy localized eigenstates approaches that of the extended states $E_\sigma$, their spatial profiles become increasingly extended. As a result, the spatial expectation values of these states tend to follow a regular pattern when plotted against the energy index. In contrast, the boundary-localized states deviate from this trend and appear as outliers in the otherwise smooth variation of spatial expectation values. Shown in Fig. 2(b), the regular progression of spatial expectation values is evident, with boundary-localized states appearing as outliers.

The classification of states in the mosaic Wannier-Stark model remains stable across various system sizes. When the system length does not match an integer multiple of the modulation period, residual sites at the boundaries lead to the emergence of additional low-energy states. These states form symmetric pairs and are strongly localized near the lattice edges. Their appearance depends on the remainder $m = N \bmod M$, and they can be reliably identified from bulk modes using combined spectral and spatial diagnostics.

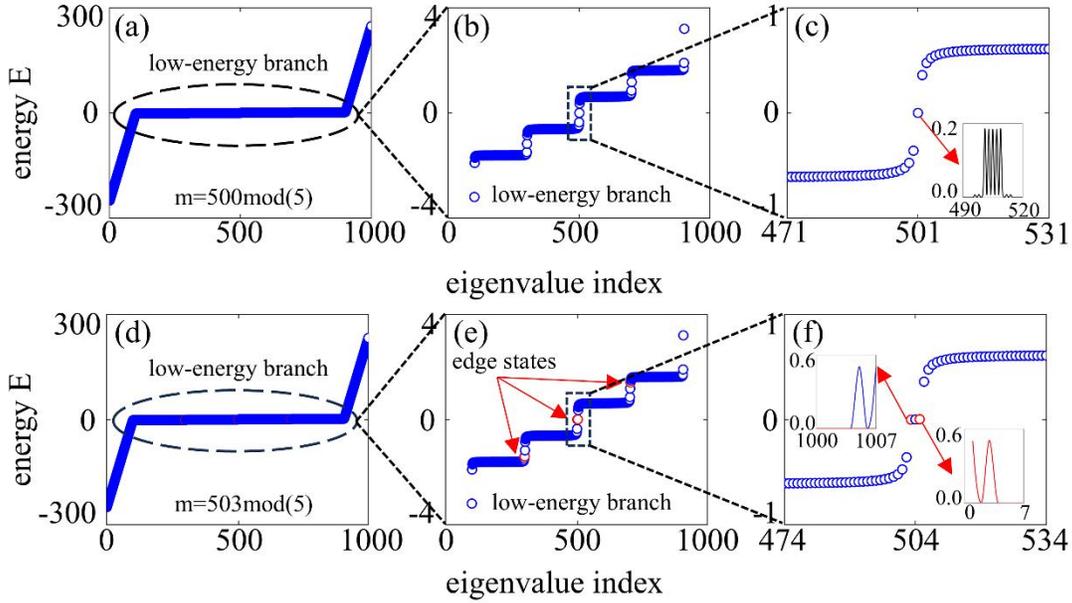

FIG. 3. Under the parameters $M = 5$ and $F/t = 0.5$ with open boundary conditions, panels (a) and (d) display the energy spectra for lattices with sizes $N = 500$ and $N = 503$, respectively. Panels (b) and (e) provide magnified views of the low-energy branches corresponding to (a) and (d), where transition energies are marked by anomalous points. These transition energies divide the energy plateaus displayed in the spectra. Panels (c) and (f) show a detailed zoom-in of one pair of anomalous points. These anomalous points, clearly marked in the figure, deviate from the regular spectral

progression and signal the onset of localization behavior. The insets in (c) and (f) display the corresponding wavefunctions, illustrating weakly localized and boundary-localized states.

We next examine the changes in the energy spectrum. The numerically obtained energy spectra for $M = 5$ with $N = 500$ and $N = 503$ under $F/t = 0.5$ confirm that the overall eigenvalue structure is largely unaffected by the residual sites, as illustrated in Fig. 3(a) and (d). We observe that the low-energy wavefunctions cluster in four narrow regions approaching the extended-state energy $E_\sigma$. Among them, the boundary-localized states are located farthest from the extended states and show stronger localization than nearby states [see Fig. 3(b) and (e)]. These observations support our assumption regarding the localization behavior in the low-energy regime.

In addition, we compare the localization differences of boundary states for $N = 503$ ($m = 3$) and $N = 500$ ($m = 0$) with low-energy states. For $N = 500$, the most localized low-energy state exhibits a wavefunction concentrated near the center of the chain, consistent with our theoretical prediction regarding the position and localization strength of boundary states. This behavior is corroborated by the inset panels in Fig. 3(c) and (f).

Overall, the analysis confirms that residual sites induce only localized modifications in the low-energy part of the spectrum. The global distribution of eigenvalues remains nearly unchanged. However, the residual sites generate additional boundary-localized states, whose spectral positions and spatial profiles are clearly distinguishable. These results provide a clearer picture of how finite-size effects, especially those related to residual sites, affect the spectral and localization properties of the mosaic Wannier-Stark model.

In realistic physical settings, perfect isolation of the system is rarely achievable. Coupling to the environment inevitably introduces energy exchange in the form of effective gain and loss, which plays a crucial role in the dynamics of open systems. To account for such effects, we consider a non-Hermitian extension of the mosaic

Wannier-Stark lattice by incorporating a complex on-site potential.

Specifically, we modify the Stark potential by adding an imaginary component and take

$$V_n = \begin{cases} Fn + i\gamma, & n = 0, \pm M, \pm 2M, \pm 3M, \ldots \\ 0, & \text{otherwise}, \end{cases} \quad (7)$$

where $\gamma \in \mathbb{R}$ represents the strength of the gain and loss, and the $F$ is the amplitude of the applied linear potential as before. This modification leads to a non-Hermitian Hamiltonian of the form

$$H = \sum_n t(c_n^\dagger c_{n+1} + \text{H.c.}) + \sum_n V_n c_n^\dagger c_{n+1}, \quad (8)$$

where all terms follow the same definitions as in the Hermitian case. Including a non-Hermitian term violates energy conservation and induces distinct spectral modifications. In the following, we systematically explore how the gain/loss parameter $\gamma$ affects the spectral structure, localization properties, and the emergence of boundary modes in the presence of periodic Stark modulation.

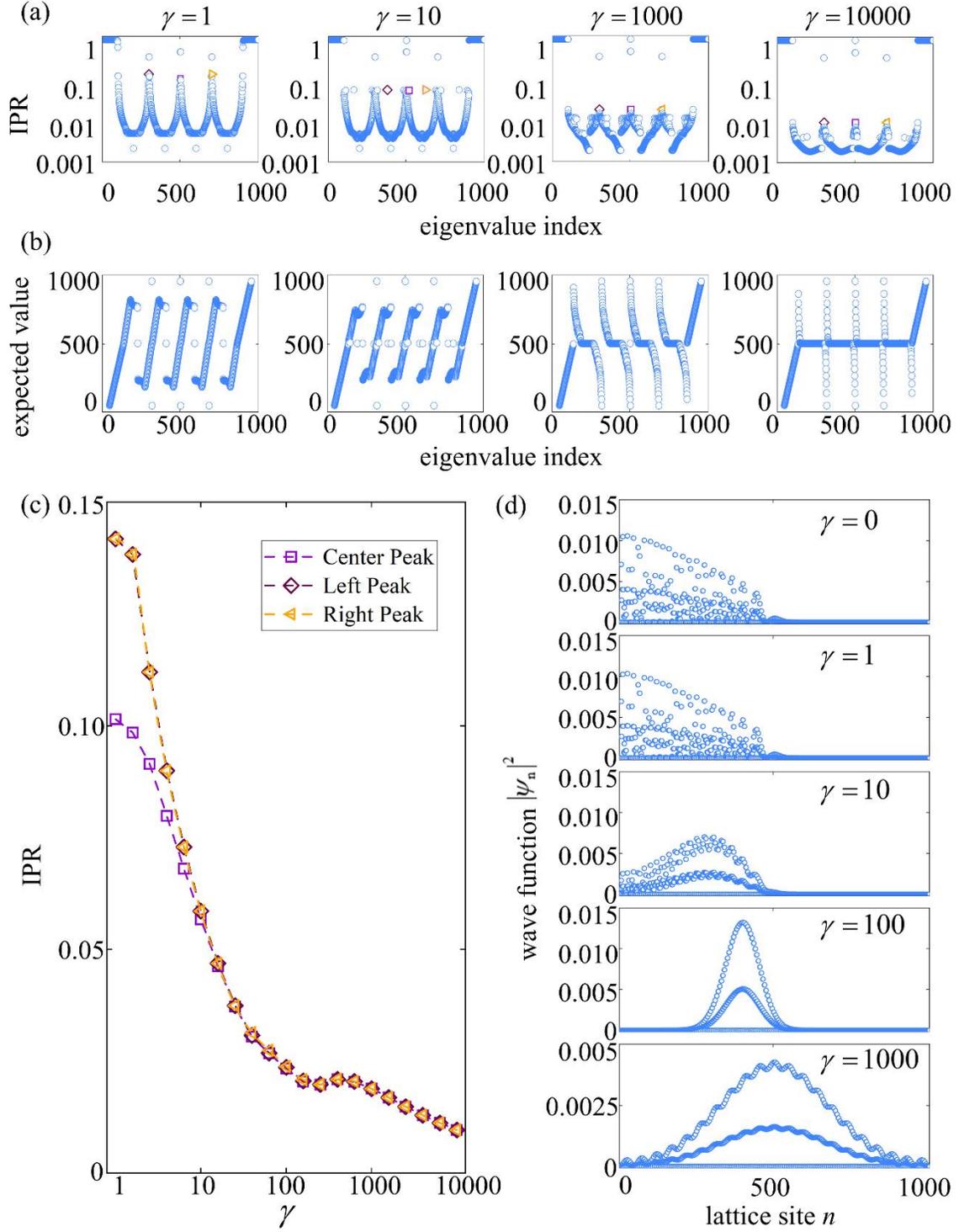

FIG. 4. For a lattice of size $N = 503$ with $M = 5$ and $F/t = 0.5$ under open boundary conditions: (a) IPRs of wavefunctions for $\gamma = $ 1, 10, 1000, 10000; (b) corresponding spatial expectation values; (c) IPRs of three representative low-energy peaks as functions of $\gamma$; (d) Spatial

profiles of the squared wavefunction amplitudes at fixed energy $E = -1.676t$ for different values of $\gamma$ on a lattice of size $N = 500$. The wavefunction exhibits weak initial localization over hundreds of lattice sites and gradually transitions to an extended state as $\gamma$ increases.

The impact of gain and loss on localization is assessed through the IPR and spatial expectation values. The gain/loss parameter $\gamma$ spans four orders of magnitude (1 to 10,000), with the corresponding spectral behavior illustrated in Fig. 4(a, b). Across all values of $\gamma$, the four isolated extended states consistently yield minimal IPR values. In contrast, the weakly localized low-energy states exhibit gradually decreasing IPRs as $\gamma$ increases, eventually aligning with the extended-state values. This trend reflects a continuous delocalization process. To visualize this effect, three representative low-energy peaks (colored points) are selected in Fig. 4(a), and their IPR evolution is presented in Fig. 4(c). The IPRs decrease monotonically, indicating progressive spatial delocalization within the low-energy branch.

We observe that the spatial expectation values are presented in Fig. 4(b). Unlike the structured profile observed in Fig. 2(b), where weakly localized states followed a regular energy-dependent pattern, the low-energy states under gain/loss gradually shift toward the center of the lattice. This spatial convergence supports our interpretation of delocalization. Our conclusion suggests that weakly localized low-energy states evolve into extended states as the gain/loss strength increases, and the data in Figs. 4(a), (b), and (c) serve to validate this hypothesis.

To provide a clearer view of the extended behavior observed in Fig. 4(a), (b), and (c), we examine a representative eigenstate at fixed energy. As shown in Fig. 4(d), we choose $E = -1.676t$ and plot the squared wavefunction amplitudes for different gain/loss parameters $\gamma =$ 1, 10, 1000, 10000 on a lattice of size $N = 500$. For $\gamma = 0$, the wavefunction is weakly confined and spreads over several hundred sites. As $\gamma$ increases, the distribution gradually becomes broader, and its peak moves closer to the center of the chain. At intermediate values such as $\gamma = 100$, the state takes on a more symmetric and wider shape. For large $\gamma$, the wavefunction extends across nearly the entire system. This smooth evolution in spatial structure illustrates the effect of non-Hermiticity on wavefunction shape. The transition from edge-localized to center-localized, and eventually to fully extended states, aligns with the IPR and spatial expectation results in Fig. 4(a–c).

Our results demonstrate that the introduction of non-Hermiticity through gain and loss drives a continuous delocalization transition in the low-energy sector. Weakly localized states gradually evolve into extended states, as evidenced by decreasing IPRs, shifting spatial expectation values, and broadening wavefunction profiles. This crossover reflects a smooth reshaping of the localization structure and highlights the role of non-Hermitian modulation in controlling spatial distributions. The consistency between IPR, spatial expectation values, and wavefunction amplitudes across a wide range of non-Hermitian strengths reinforces our interpretation of gain/loss-induced delocalization.

In summary, we explore the spectral and localization properties of the mosaic Wannier-Stark lattice. When the system length is incommensurate with the modulation period, residual sites at the lattice edges give rise to the emergence of additional low-energy boundary-localized states. These states are sharply confined near the boundaries and are clearly distinguishable from both the weakly localized bulk modes and the isolated extended states of the infinite system. By employing the IPR and spatial expectation values, we present an effective diagnostic for identifying and characterizing the localization features of these states. Furthermore, we extend the model to the non-Hermitian regime by introducing complex on-site potentials to simulate gain and loss. Our results indicate that the low-energy localized states undergo progressive delocalization as the non-Hermitian strength increases, exhibiting a smooth crossover toward spatially extended behavior. These findings offer new insights into boundary-induced localization phenomena and the interplay between finite-size effects and non-Hermitian physics.

**Acknowledgements**

The authors thank for the support by National Natural Science Foundation of China under (Grant 12404365) and useful discussions with Dr. Tong Liu.